\documentclass[12pt,aps,preprint,nofootinbib,superscriptaddress,nobalancelastpage]{revtex4}

\usepackage{hyperref}
\usepackage{epsfig}
\usepackage{amssymb}
\usepackage{amsmath}
\usepackage{amsfonts}
\usepackage{graphics}
\usepackage{cancel}
\usepackage{bbm}
\usepackage{mathrsfs}
\usepackage{slashed}
\usepackage{color}
\usepackage{subfigure}

\newcommand{\Ref}{Ref.}

\newcommand{\eq}{Eq.}

\newcommand{\fig}{Fig.}

\newcommand{\bea}{\begin{eqnarray}}
\newcommand{\eea}{\end{eqnarray}}

\newcommand{\optbar}[1]{\shortstack{{\tiny (\rule[.4ex]{1em}{.1mm})}
  \\ [-.7ex] $#1$}}		% Still another (---)

\newcommand{\be}{\begin{equation}}
\newcommand{\ee}{\end{equation}}

\newcommand{\ba}{\begin{array}}
\newcommand{\ea}{\end{array}}

\newcommand{\ie}{\emph{i.e.}}

\allowdisplaybreaks

\begin{document}
\title{Optimization of neutrino oscillation facilities for large $\theta_{13}$}

\author{Pilar Coloma}
\email[]{p.coloma@uam.es}
\affiliation{Departamento de F\'isica Te\'orica \& Instituto de F\'isica Te\'orica, Universidad Aut\'onoma de Madrid,
28049 Madrid, Spain}

\author{\mbox{Enrique~Fernandez-Martinez}}
\email[]{enfmarti@cern.ch}
\affiliation{CERN Physics Department, Theory Division, CH-1211 Geneva 23, Switzerland}

\begin{abstract}
Up to now, future neutrino beam experiments have been designed and optimized in order to look for CP violation, $\theta_{13}$ and the mass hierarchy under the conservative assumption that $\theta_{13}$ is very small. However, the recent results from T2K and MINOS favor a $\theta_{13}$ which could be as large as $8^\circ$. In this work, we propose a re-optimization for neutrino beam experiments in case this hint is confirmed. By switching from the first to the second oscillation peak, we find that the CP discovery potential of future oscillation experiments would not only be enhanced, but it would also be less affected by systematic uncertainties. In order to illustrate the effect, we present our results for a Super-Beam experiment, comparing the results obtained at the first and the second oscillation peaks for several values of the systematic errors. We also study its combination with a $\beta$-beam facility and show that the synergy between both experiments would also be enhanced due to the larger $L/E$. Moreover, the increased matter effects at the longer baseline also significantly improve the sensitivity to the mass hierarchy. 
\end{abstract}


\preprint{CERN-PH-TH/2011-255, EURONU-WP6-11-41, IFT-UAM/CSIC-11-72}

\maketitle

%%%%%%%%%%%%%%
\section{Introduction}

The discovery of neutrino oscillations demands some extension of the Standard Model of particle physics leading to neutrino masses and flavour mixing in the lepton sector. Despite the progress in our understanding of neutrino physics over the last years, we remain ignorant of the mechanism behind neutrino masses and the full pattern of masses and mixings is, as yet, incomplete. Two distinct regimes have been observed (see \Ref~\cite{Schwetz:2011zk} for a recent global fit). Atmospheric neutrino data as well as long baseline experiments with neutrino beams from accelerators require a mass splitting of $\Delta m^2_{31} = 2.5 \cdot 10^{-3}$ eV$^2$ and a nearly maximal mixing angle $\theta_{23} \sim 45^\circ$. Solar and reactor neutrino data, on the other hand, show oscillations with much longer periods, corresponding to a smaller splitting of $\Delta m^2_{21} = 7.6 \cdot 10^{-5}$ eV$^2$ and a non-maximal, although large, mixing angle, $\theta_{12} = 34^\circ$. The ordering of the neutrino masses, \ie{} whether a normal or inverted hierarchy is realized in nature, as well as the absolute neutrino mass scale remain unknown. Similarly, the third mixing angle, $\theta_{13}$, and the existence of leptonic CP violation have not yet been probed. 

New results from the T2K experiment~\cite{Abe:2011sj}, MINOS~\cite{Adamson:2011qu} and Double-CHOOZ~\cite{Kuze:2011ic} favour large values of $\theta_{13}$, saturating the present constraints. A global fit to present data yields a preference for non-zero $\theta_{13}$ at $\sim 3 \sigma$ with a best fit at $\sin^2 2\theta_{13} = 0.051$ $(0.063)$ for normal (inverted) hierarchy~\cite{Schwetz:2011zk}, see also \cite{Fogli:2011qn,Machado:2011ar}. If confirmed with larger statistics and by the ongoing reactor searches~\cite{Ardellier:2006mn,Guo:2007ug}, this would imply that our ability to probe for leptonic CP violation and determine the neutrino mass hierarchy are closer at hand than we dared hope for. 
In such an event, we should evaluate the optimization of future oscillation facilities to measure these two observables. Indeed, most neutrino oscillation experiment proposals choose their energy and baseline so as to observe the $\nu_e \to \nu_\mu$ oscillations of neutrinos and antineutrinos (or its T conjugates) at the first maximum of the ``atmospheric'' $\Delta m^2_{31}$ oscillation. The vacuum oscillation probability for this channel, expanded up to second order\footnote{For large $\theta_{13}$, a higher order expansion in this parameter, as provided in Ref.~\cite{Asano:2011nj}, would better reproduce the exact results. We will only use this approximation as a guideline and use the exact probability for all numerical simulations.} in $\theta_{13}$ and the ``solar'' $\Delta m^2_{21}$ splitting reads~\cite{Cervera:2000kp}: 

\begin{eqnarray}
P^{\pm}_{e \mu} \equiv P(\optbar{\nu_e} \to \optbar{\nu_\mu})  & = &
s_{23}^2 \, \sin^2 2 \theta_{13} \, \sin^2 \left ( \frac{\Delta_{31} \, L}{2} \right ) +
c_{23}^2 \, \sin^2 2 \theta_{12} \, \sin^2 \left( \frac{ \Delta_{21} \, L}{2} \right ) \nonumber \\
& + & \tilde J \, \cos \left ( \pm \delta - \frac{\Delta_{31} \, L}{2} \right ) \;
\sin \left(\frac{ \Delta_{21} \, L}{2}\right ) \sin \left ( \frac{\Delta_{31} \, L}{2} \right ) \, ,
\label{eq:vacexpand}
\end{eqnarray}
where the upper/lower sign in the formula refers to neutrinos/antineutrinos, $\tilde J \equiv c_{13} \, \sin 2 \theta_{12} \sin 2 \theta_{23} \sin 2 \theta_{13}$ and $\Delta_{ij} \equiv \frac{\Delta m^2_{ij} }{2 E_\nu}$. We will refer to the three terms in \eq~(\ref{eq:vacexpand}) as ``atmospheric'', ``solar'' and ``CP interference'' terms, respectively.

\begin{figure}
\begin{center}
\includegraphics[width=.48\textwidth]{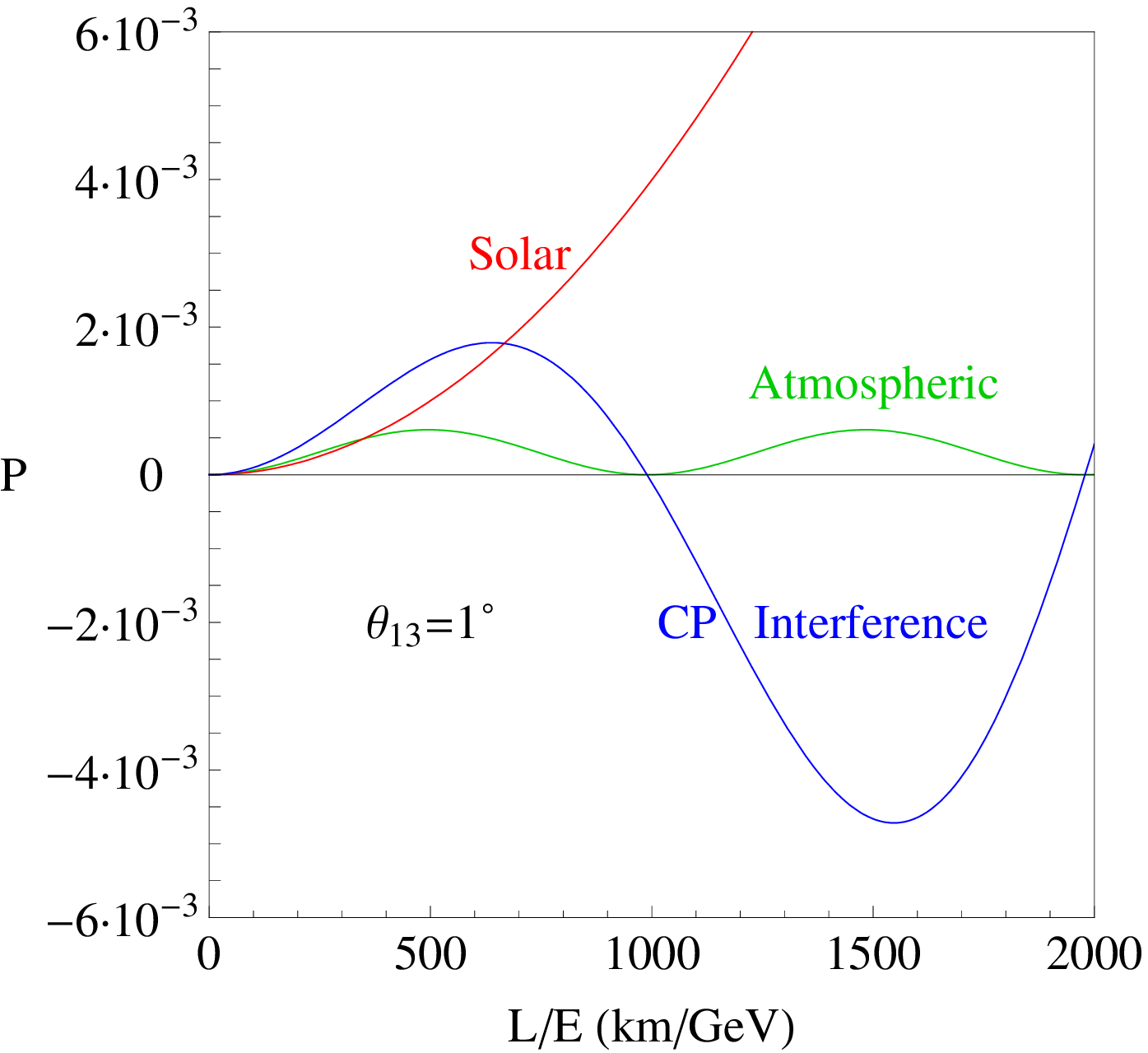}
\includegraphics[width=.48\textwidth]{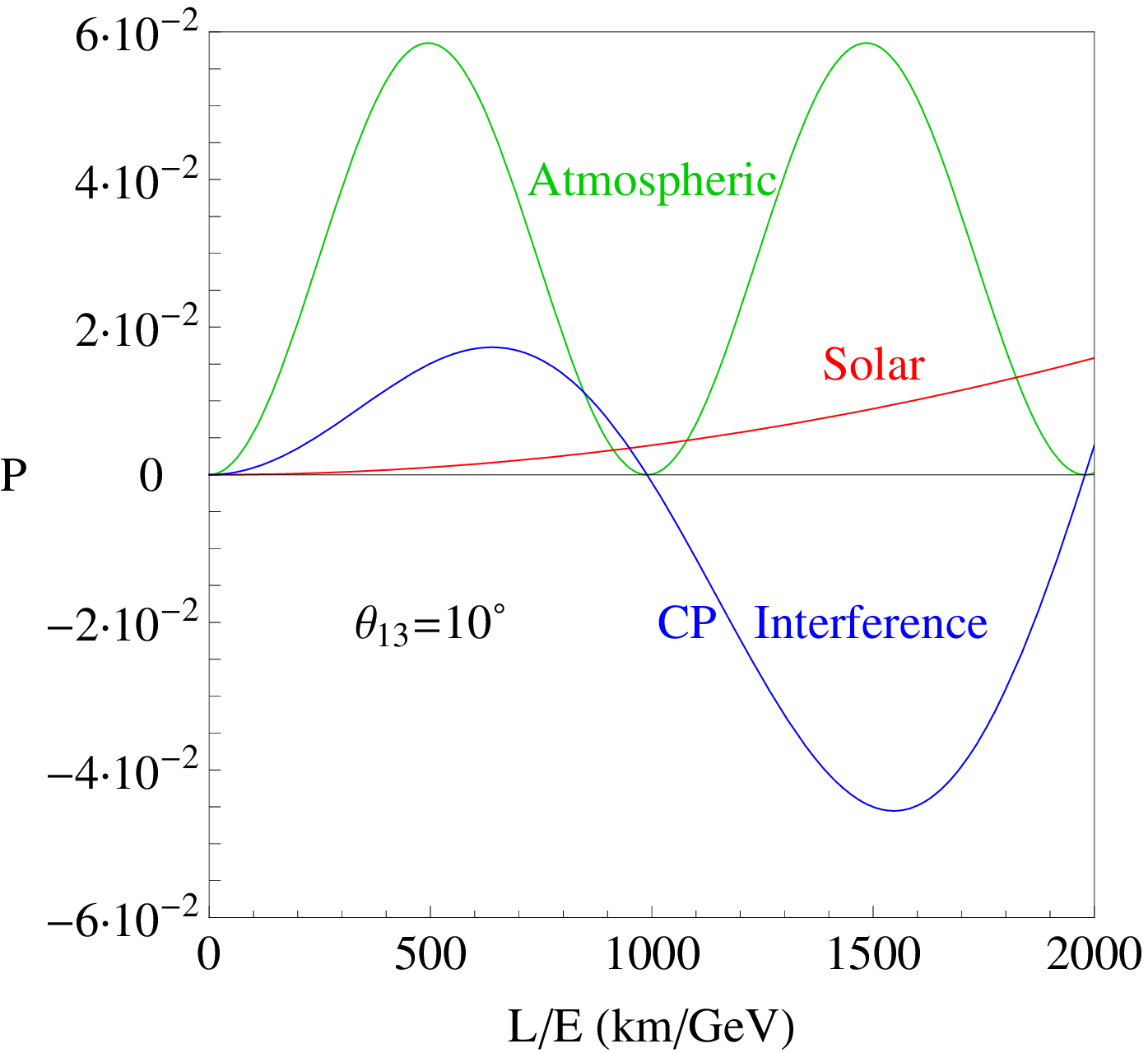} \\
\caption{Terms of the oscillation probability in vacuum as a function of $L/E$ for $\theta_{13}=1^\circ$ (left) and $\theta_{13}=10^\circ$ (right). Notice the different scales in the Y-axis between the two panels. The terms driven by the ``atmospheric'' (green) and ``solar'' (red) oscillation frequencies as well as the CP-violating interference (without the $\cos( \pm \delta - \frac{\Delta_{31} \, L}{2})$ term) between the two (blue) are shown. }
\label{fig:probs}
\end{center}
\end{figure}

In \fig~\ref{fig:probs} the three terms in \eq~(\ref{eq:vacexpand}) are depicted as a function of $L/E$. The left panel shows the case of $\theta_{13}=1^\circ$, 
while the right panel corresponds to $\theta_{13}=10^\circ$ (close to the best fit of T2K). For the CP-violating interference term only the coefficient in front of $\cos \left ( \pm \delta - \frac{\Delta_{31} \, L}{2} \right )$ has been shown. As can be seen, for $\theta_{13}=1^\circ$ the choice of the first oscillation peak is indeed very favorable for the exploration of CP violation, since the coefficient multiplying the CP-violating term is larger than either the solar or the atmospheric CP-conserving terms. On the other hand, for $\theta_{13}=10^\circ$ the first oscillation peak is dominated by the atmospheric term whereas the CP interference term is only a subleading component of the oscillation probability which could be missed unless the systematic error on the signal is kept very low. Indeed, in order to maximize the interference between the two terms, they should both be of the same order, \ie{} for large $\theta_{13}$ a longer $L/E$ would be preferable so that ``solar'' oscillations develop to a similar size: $\sin 2 \theta_{13} \sim \Delta m_{21}^2 L/(4E)$, while the first oscillation peak corresponds to $\Delta m^2_{13} L/(4E) \sim \pi/2$. Thus, these two criteria only coincide when $\sin 2 \theta_{13} \sim \pi/2  \Delta m_{21}^2/\Delta m_{31}^2 \sim 0.05$. This implies that, for ``large'' values of $\sin 2 \theta_{13}$ (currently favored by data) somewhat larger values of $L/E$ would be preferred, since they would enhance the CP-violating contribution with respect to the atmospheric one and the dependence of the CP discovery potential on the systematic error would consequently decrease. 

Here we will show how the displacement of a neutrino oscillation facility from the first to the second oscillation peak can enhance both its CP discovery potential and sensitivity to the mass hierarchy for large values of $\theta_{13}$, providing a better alternative if the present hint from T2K is confirmed by ongoing reactor searches \cite{Ardellier:2006mn,Guo:2007ug}. Moreover, as we will show, the larger $L/E$ also makes the CP discovery potential more stable against systematic uncertainties for large $\theta_{13}$, since the CP interference term will become a leading part of the oscillation probability and hence harder to hide behind systematic errors. Given that systematics errors dominate the sensitivity of Super-Beams at large $\theta_{13}$~\cite{Huber:2007em} this will be particularly desirable unless they can be controlled to a very low level. 

The idea of including information on the second oscillation maximum in combination with the data from the first peak is an old and very well studied one in the context of on-axis, wide band beam fluxes, such as the one proposed for the LBNE experiment~\cite{Diwan:2003bp,Barger:2006vy,Barger:2007yw,Agarwalla:2010nn,Huber:2010dx}, that can cover the first two oscillation peaks with their wide neutrino spectra. This combination potentially offers a strong complementarity between the lower and higher end of the neutrino spectra that could allow to solve degeneracies and increase the sensitivity of the facility~\cite{Diwan:2003bp}. However, the neutral current background from the high energy end of the LBNE spectrum tuned to the first oscillation maximum is reconstructed at low energies, thus overwhelming the sample corresponding to the second oscillation peak (see Fig~2 of Ref.~\cite{Huber:2010dx}). This renders it almost useless quantitatively~\cite{Huber:2010dx}. The idea we explore in this work is very different. We are not interested in the potential of the second oscillation peak as a complement to data at the first oscillation maximum so as to solve degeneracies and increase its sensitivity. We rather propose not to study the neutrino oscillation at the first peak, given its reduced sensitivity to both CP violation and the mass hierarchy, and focus the search with a narrow beam around the more useful second peak instead, therefore avoiding the neutral current background from the high energy tail of the spectrum that would spoil its sensitivity. Moreover, while the events observed at the second peak have a stronger dependence on leptonic CP violation and, given the longer baseline, also to the mass hierarchy than those at the first peak, they also suffer from reduced statistics, given the longer baselines or smaller energies required. Thus, to overcome the lower statistics expected, not too long baselines are preferable. It is then not surprising that the second oscillation maximum did not prove very useful in the study of Ref.~\cite{Huber:2010dx} for the long baselines associated to the LBNE setup. In this context, the proposal of studying the neutrino beam from Tokai at a detector in Korea, T2KK~\cite{Ishitsuka:2005qi,Dufour:2010vr}, is more similar to the idea discussed here, although the baseline is still much longer than the one we consider and thus, less optimal for the study of the second oscillation maximum. Moreover, the stronger matter effects found at longer baselines and higher energies modify the oscillation frequency of neutrinos and antineutrinos in different ways so that tuning both beams to the second oscillation peak becomes challenging and less optimal.

The paper is organized as follows. In Section~\ref{sec:setups} we introduce the experimental setup of the SPL Super-Beam and its companion $\beta$-Beam. In Section~\ref{sec:results} we show our results: we first show the effect on the CP discovery potential and the sensitivity to the mass hierarchy when the experiment is performed at the second oscillation peak and compare with the results obtained at the first oscillation peak; we also show how the combination with a $\beta$-Beam facility performs better at larger values of $L/E$, as well as the dependence of our results on the systematic errors. Finally, in Section~\ref{sec:summary} we summarize and draw our conclusions.

%%%%%%%%%%%%%%
\section{Setups}
\label{sec:setups}

%In order to explore whether facilities detecting neutrino beams beyond the first oscillation peak can provide increased CP discovery potential as well as less dependence on the unavoidable systematic errors, as outlined in the introduction, 

The main purpose of this work is to explore whether the detection of neutrino beams beyond the first oscillation peak can enhance their CP discovery potential. In order to address this question, we will study the specific case of facilities with rather weak matter effects, \ie{} with short baselines and low energies.  This choice is motivated by the fact that strong matter effects modify the oscillation frequency of neutrinos and antineutrinos in different ways, and the corresponding discussion and baseline optimization becomes rather complicated. Therefore, as a first step, we will take a well-studied low energy and short baseline facility, normally taken to be close to the first oscillation peak: the SPL Super-Beam. This facility is commonly taken in combination with a $\gamma = 100$ $\beta$-Beam. We will also show how such combination can be more effective for large $\theta_{13}$ at larger $L/E$.

\subsection*{The Super-Beam}

\begin{figure}[htp]
\centering
\subfigure[~SB flux in $\nu$ mode]{
          \includegraphics[scale=0.35]{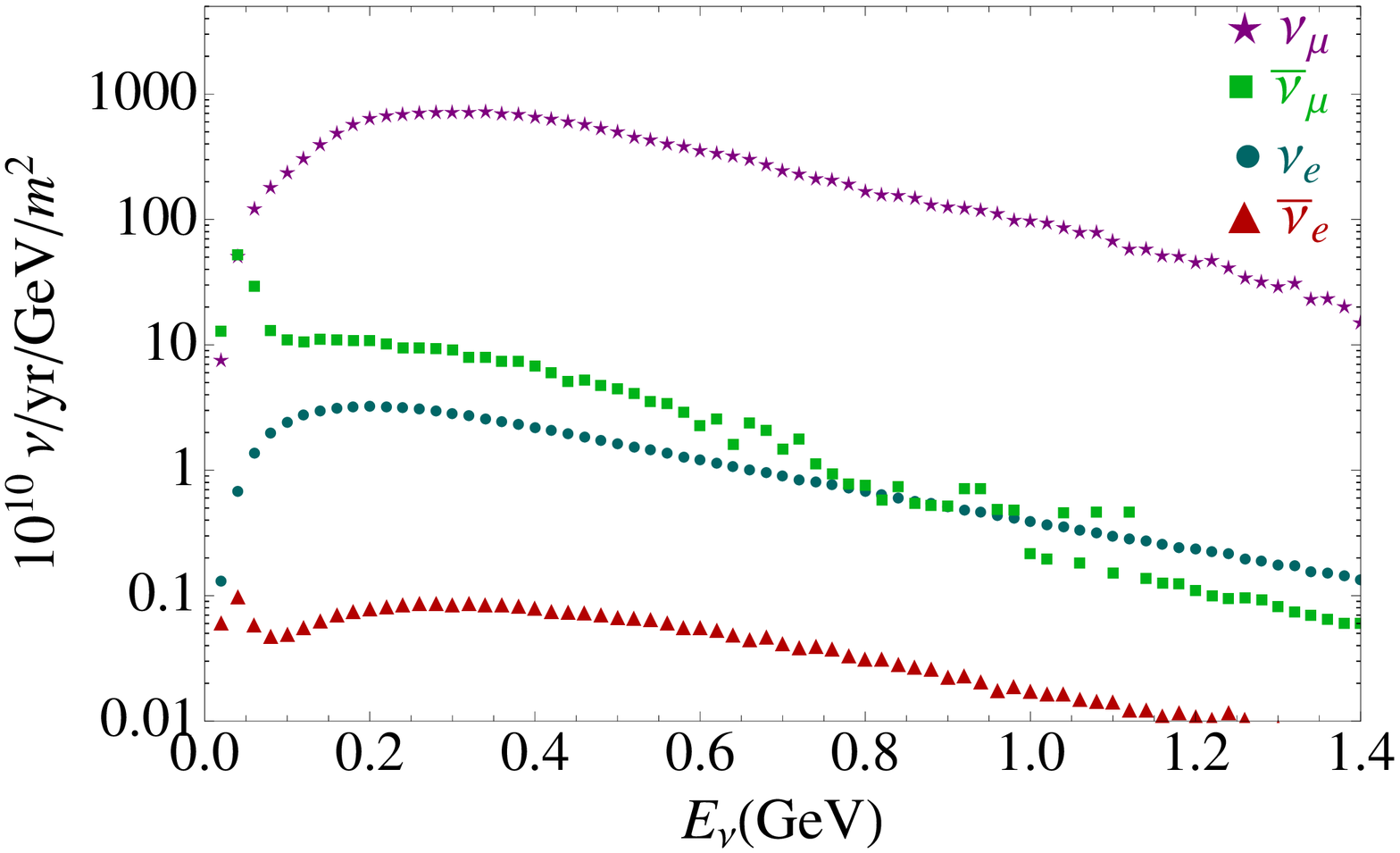}\label{fig:SBflux}}
\subfigure[~$\nu_e$ and $\bar\nu_e$ $\beta$-Beam fluxes]{
          \includegraphics[scale=0.35]{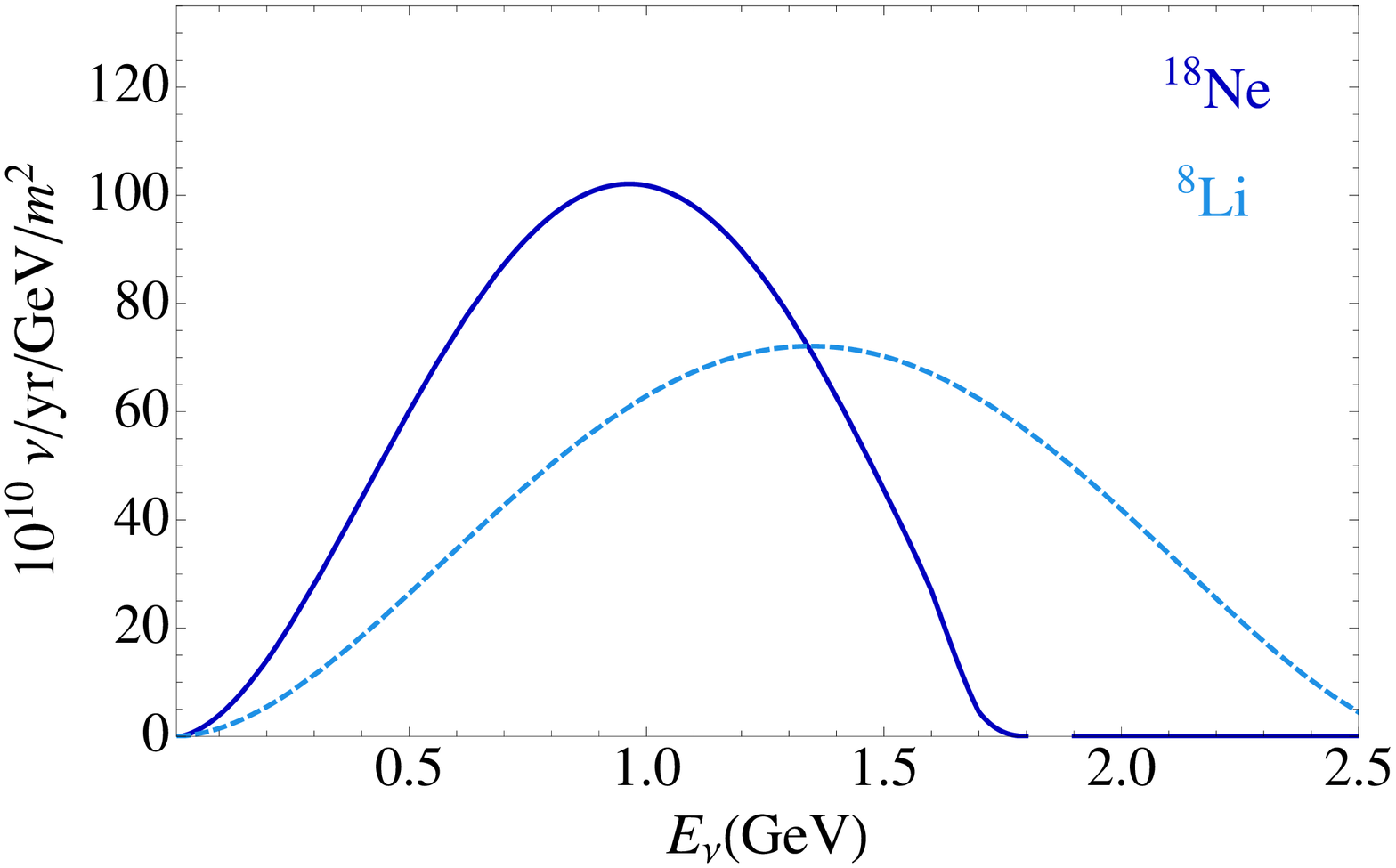}\label{fig:BBflux}}\\
\caption[Beam content]{Fluxes for the two facilities under study, measured at 100 km from the source, as a function of the neutrino energy: a) Super-Beam beam composition for positive horn focusing from Ref.~\cite{Longhin:2011hn}, in logarithmic scale; b) $\nu_e$ and $\bar\nu_e$ flux for the $\beta$-Beam produced from $^{18}\textrm{Ne}$ and $^8\textrm{Li}$ boosted to $\gamma=250$ and 100, for $0.44 \times 10^{18}$ and $2.8\times 10^{18}$ useful ion decays per year, respectively.}
\label{fig:fluxes}
\end{figure}

A Super-Beam is a conventional neutrino beam driven by a proton driver with a beam power in the range 2-5 MW~\cite{Richter:2000pu,Diwan:2004bt}. At these facilities, neutrinos are produced from the decays of pions and kaons. Therefore, together with the desired $\nu_\mu$ ($\bar\nu_\mu$) flux a small but unavoidable mixture of $\bar\nu_\mu$, $\nu_e$ and $\bar\nu_e$ will also be produced (see Fig.~\ref{fig:SBflux}). The main channels available at this kind of experiments are the $\nu_\mu \rightarrow \nu_\mu$ and $\nu_\mu\rightarrow \nu_e$ channels. The disappearance channel would mainly be useful to measure the atmospheric parameters while the appearance channel is the one which provides sensitivity to CP violation. The main advantage of Super-Beam facilities is that they profit from a well established production technology. Their main drawback, on the other hand, is the intrinsic contamination of the beam, which affects the sensitivity to $\nu_\mu \rightarrow \nu_e$ oscillations. A further limitation of this kind of experiment is the flux uncertainty, which affects both signal and background predictions and constitutes an additional source of systematic errors. 

%%%%%%%%%%%%%%%%%%%%
%%%%%%%%%%%%%%%%%%%%

Here we will study the SPL Super-Beam, designed for the CERN complex and originally conceived for the CERN to Fr\'ejus baseline of 130 km~\cite{GomezCadenas:2001eu,Donini:2004hu,Campagne:2004wt,Donini:2004iv,Donini:2005db,Campagne:2006yx} but with aimed at a longer baseline of 650 km, matching the CERN to Canfrac distance, which is better suited to study the second oscillation peak. The Super-Beam considered here can be regarded as an optimization of the SPL (with reduced beam contamination and a broader peak, see Ref.~\cite{Longhin:2011hn} for a detailed discussion) and is originated from 4.5 GeV protons, inciding on the target at a rate of $0.56\times10^{23}$ Protons on Target (PoT) per year. In order to overcome the smaller antineutrino cross section as well as the lower flux when the experiment is run with negative horn focusing, the experiment is assumed to run for 8 years in antineutrino mode (negative horn focusing) and only 2 years in neutrino mode (positive horn focusing), so as to guarantee a similar number of events for both polarities at the detector. The original proposal for the detector for this beam is a Mton (440 kton fiducial) Water \v{C}erenkov (WC) detector~\cite{deBellefon:2006vq}. As a reference, we will also compare all the results with the performance of the T2HK setup~\cite{Abe:2011ts}. We have however increased the proposed beam power to 4 MW and we consider a fiducial detector mass of 440 kton as well as 5 years running in neutrino and antineutrino mode so as to make a comparison on equal footing with the SPL scenario. The detector response in terms of efficiencies and backgrounds for both the SPL and T2HK has been taken from Ref.~\cite{Campagne:2006yx}. The treatment of the systematic errors is also somewhat different in Ref.~\cite{Abe:2011ts}. However, we decided to include the systematic errors in the same way as for the SPL setup so as to allow a direct comparison on the relative impact of systematics at each facility.

\subsection*{The $\beta$-Beam}

At a $\beta$-Beam, neutrinos are produced from the decay of $\beta$-unstable ions in the straight sections of a storage ring aiming to a far detector. In the original proposal~\cite{Zucchelli:2002sa}, neutrinos and antineutrinos are produced from the $\beta$-decay of $^6\textrm{He}$ and $^{18}\textrm{Ne}$ boosted to $\gamma=100$. Therefore, the flux is only composed of electron (anti)neutrinos and its composition is known precisely. The main channel that can be observed at this facility is the ``golden'' channel~\cite{Cervera:2000kp} ($\nu_e\rightarrow \nu_\mu$ and its CP conjugate), which would be measured through the observation of $\mu^\pm$ at the detector. 

As for the SPL Super-Beam, the proposal for the detector is a Mton-scale WC placed at the Fr\'ejus site, $L=130$ km from CERN. Detailed analyses of the physics performance of this setup can be found in Refs. \cite{Mezzetto:2003ub,Mezzetto:2004gs,Donini:2004hu,Donini:2004iv,Donini:2005rn,Huber:2005jk,Campagne:2006yx, FernandezMartinez:2009hb}. Numerous modifications of this basic setup have been studied \cite{Agarwalla:2005we,Donini:2006tt,Volpe:2006in,Agarwalla:2006vf,Donini:2007qt,Jansson:2007nm,Agarwalla:2007ai,Coloma:2007nn, Meloni:2008it,Agarwalla:2008gf,Agarwalla:2008ti,Winter:2008cn,Winter:2008dj,Choubey:2009ks,Coloma:2010wa}, most of them being different combinations of two basic ingredients: the possibility of accelerating the ions to higher $\gamma$ factors \cite{BurguetCastell:2003vv,BurguetCastell:2005pa}, thus increasing the flux and the statistics at the detector, and the possibility of considering the decay of different ions to produce the neutrino beam. In particular $^8$B and $^8$Li
have been proposed as alternatives to $^{18}$Ne and $^{6}$He respectively~\cite{Rubbia:2006pi,Donini:2006dx,Rubbia:2006zv}. 
Among these variations, the higher $\gamma$ factors are particularly desirable since both the neutrino energy and the collimation of the beam increase with $\gamma$, thus increasing the statistics at the detector. Indeed, the $\beta$-Beam performance seems now to be mainly statistically limited due to technical difficulties related to ion production and acceleration, specially compared with the recently optimized Super-Beam fluxes. The ``nominal'' neutrino fluxes typically assumed in the literature for this facility are obtained from $1.1\times10^{18}$ and $2.9\times 10^{18}$ useful ion decays per year for $^{18}\textrm{Ne}$ and $^6\textrm{He}$, respectively. For the $^8$B and $^8$Li alternatives its production rates have been studied in much less detail. While $^8$Li can also be produced with ISOLDE techniques, alternatives such as the ``ionization cooling'' technique proposed in Ref.~\cite{Rubbia:2006pi} need to be explored for $^8$B. In any case, the production of $^8$Li also seems easier than that of $^8$B through ionization cooling. For this reason we will not consider $^8$B ions in any of the setups studied and we will assume that $2.9\times 10^{18}$ useful $^8$Li decays per year are achievable for $\gamma=100$. When considering $\gamma$ factors larger than the usual $\gamma=100$ we will also reduce the ion flux by a factor $100/\gamma$ so as to take into account the boosted ion lifetime (that implies a reduced flux if the size of the decay ring is kept as for the $\gamma = 100$ option).

The main advantage of the $\beta$-Beam scenario over the Super-Beam alternative is, therefore, its purer neutrino flux with a lower expected systematic uncertainty and no beam contamination. On the other hand, the $\beta$-Beam technology is much more speculative as compared to the Super-Beam and achieving the targeted beam intensities seems challenging. When compared to the newly optimized Super-Beam fluxes the $\beta$-Beam shows a significant statistics limitation (Fig.~\ref{fig:BBflux}). Moreover, the background from atmospheric neutrinos is one of the limiting factors for the $\beta$-Beam sensitivity. This background can be reduced by imposing angular cuts in the direction of the beam. Indeed, even if the direction of the incoming neutrino cannot be measured, it is increasingly correlated with the direction of the detected muon at higher energies. This situation is depicted in Fig.~5 of Ref.~\cite{BurguetCastell:2005pa} for neutrinos from $^{18}$Ne and $^6$He with $\gamma$ factors of $120$, $150$ and $350$. In particular the energy spectrum of neutrinos from $^8$Li with $\gamma=100$ is very similar to $^{18}$Ne at $\gamma=350$ since their decay energy is precisely $\sim 3.5$ times larger. As can be seen from the figure, the mean angle between the muon and the incoming neutrino is much smaller in the $\gamma = 350$ scenario. An angular cut requiring $90 \%$ of the efficiency was applied in \cite{BurguetCastell:2005pa} and is included in the migration matrices extracted from that reference and used here.  In order to estimate the atmospheric neutrino background surviving this cut, we have evaluated the expected number of oscillated atmospheric muon neutrinos using the new flux from \cite{Honda:2011nf} for the Fr\'ejus site that arrived within a solid angle $\sim \sqrt{1/E\textrm{(GeV)}}$ from the beam direction for the different energy bins considered. This corresponds to a conservative estimate of the typical scattering angle between a parent neutrino with energy $E$ and the final state charged lepton. The results agree well with the backgrounds rates quoted by \cite{BurguetCastell:2005pa}.

Previous analyses showed that, in order to reduce the remaining atmospheric neutrino rate below the neutral current beam-induced background for the standard setup, the decaying ions must be accumulated in very small bunches so as to achieve a $10^{-4}$ suppression factor of the background~\cite{BurguetCastell:2005pa,Mezzetto:2005ae,FernandezMartinez:2009hb} through a timing cut. Recent studies show that such stringent suppression factors are rather challenging for the $\beta$-Beam\footnote{This background should also be considered, in principle, for the Super-Beam since the neutrino flux lies in the same energy range. However, for the SPL much more stringent duty cycles (around $2\times 10^{-4}$) are technically feasible and therefore this background would be negligible~\cite{Campagne:2006yx}. }, specially in combination with achieving the ``nominal'' neutrino fluxes, and they are unlikely to get better than the $10^{-2}$ level, see Ref.~\cite{Hansen:2011us} for a recent discussion. We will therefore not neglect the atmospheric background when simulating the $\beta$-Beam setup, as is usually done in the literature, and just assume a $10^{-2}$ suppression factor in the simulated background in combination with the ``nominal'' fluxes.  

%We have considered the same ions and fluxes as in the original proposal for the $\beta$-Beam aimed at $130$ km, which match the first oscillation peak at this baseline. 
As already mentioned, the fluxes produced from the decays of $^{18}\textrm{Ne}$ and $^6\textrm{He}$ ions boosted at $\gamma=100$ perfectly match the $1^\textrm{st}$ oscillation peak at 130 km.
However, since the $\beta$-Beam is mainly afflicted by statistical limitations, going to the second peak would necessarily imply a huge increase in the number of useful ion decays per year in order to achieve reasonable statistics at the detector. Therefore, in this case we have decided to try to keep the oscillation as closer to the first peak as possible. The first oscillation peak at this baseline takes place around 1.2 GeV, which could be reached for $^6$He and $^{18}$Ne boosted to $\gamma\sim 350$~\cite{BurguetCastell:2003vv, BurguetCastell:2005pa}. However, the maximum boost factors attainable at the SPS for $^6$He and $^{18}$Ne are $\gamma=150$ and $250$, respectively. For these values the neutrino flux would be near the first oscillation peak, but the antineutrino flux would be too far away from it. Thus, we have used $^8$Li boosted to $\gamma=100$ in this case instead. As it can be seen from Fig.~\ref{fig:BBflux}, the antineutrino flux in this case is centered around 1.3 GeV therefore matching perfectly the first oscillation peak for $L=650$ km. 

\subsection*{Detector details}

As it can be seen from Fig.~\ref{fig:fluxes}, all the beams that will be considered in this paper have their peak at low energies, below 1.5 GeV or so. The optimal detector should have a very good energy resolution and good reconstruction efficiencies for neutrino events in the QE regime. It should also be able to detect and correctly identify muons (in the case of a $\beta$-Beam) and electrons (in the case of a Super-Beam). Among the different detectors which satisfy these requirements, the WC presents an additional advantage: it can be built on the Mton scale, which would be of great help in overcoming the statistical limitations of setups with larger $L/E$ ratios, such as the ones considered here. A 1 Mton WC detector (440 kton fiducial) near the CERN accelerator complex could be hosted either at Fr\'ejus~\cite{deBellefon:2006vq}, (at a distance of $L=130$ km from CERN) or at Canfranc (at a distance of $L=650$ km). Both of these options are considered inside the LAGUNA (Large Apparatus for Grand Unification and Neutrino Astrophysics) design study as possible sites to allocate a very massive deep-underground particle detector~\cite{laguna,Autiero:2007zj}. A third possibility, also considered within LAGUNA, is Umbria (Italy), at a distance of $L=665$ km from CERN. However, an entire new laboratory should be built in this case~\cite{Rubbia:2010fm}, approximately 10 km away from the present Gran Sasso Laboratory. 

In order to simulate the WC response when exposed to our Super-Beam we have followed Ref.~\cite{Campagne:2006yx}, where the response of this kind of detector exposed to the SPL beam was studied in detail. The WC detector response when exposed to a $\beta$-Beam flux for different $\gamma$ factors was studied in \cite{BurguetCastell:2005pa} and the migration matrices provided there have been used to simulate the signal and beam-induced backgrounds at the WC detector. The cross sections used have been taken from Ref.~\cite{Lipari:1994pz}. We have verified that with these assumptions the SPL beam, when observed at a 650 km baseline, is mainly at the second oscillation maximum. Indeed, while the first peak is at 1.3 GeV, the signal window only spans up to 1 GeV. Even if all events are thus beyond the first peak and have the stronger dependence on CP violation that we sought, it is true that the last three energy bins are closer to the first peak than to the second. We have checked that the impact of this three energy bins in the improvement of the CP discovery potential is minimal. They can, however, prove useful in the determination of the mass hierarchy, given their higher energy and consequently stronger matter effects 

%%%%%%%%%%%%%%
\section{Results}
\label{sec:results}

In this section we will show how the performance of the SPL Super-Beam at probing the mass hierarchy and leptonic CP-violation can improve for large $\theta_{13}$ when moving from the first to the second oscillation peak, given the stronger matter effects at longer baselines. For the numerical simulations of this section the following values of the oscillation parameters were assumed: $\theta_{12} = 34^\circ$, $\theta_{23} = 42^\circ$, $\Delta m^2_{21} = 7.6 \cdot 10^{-5}$ eV$^2$ and $\Delta m^2_{31} = 2.5 \cdot 10^{-3}$ eV$^2$. All these parameters were left free in the fit and marginalized over with the following gaussian $1 \sigma$ error priors: $3 \%$ for $\theta_{12}$, $8 \%$ for $\theta_{23}$, $2.5 \%$ for $\Delta m^2_{21}$ and $4 \%$ for $\Delta m^2_{31}$, roughly corresponding to present day uncertainties (see, for instance, Ref.~\cite{Schwetz:2011zk} for a recent global fit). A conservative $5\%$ error over the PREM density profile~\cite{prem} was also considered. A $5 \%$ systematic error has been assumed for the signal channels and a $10 \%$ for the background except for Fig.~\ref{fig:sys}, where the effect of increasing and reducing the systematics by a factor 2 was explored. These systematics are fully correlated among the different energy bins of a particular channel and uncorrelated between the neutrino/antineutrino channels. While some sources of systematics, such as the fiducial volume of the detector and -to some extent- the neutrino cross sections, will be correlated between the neutrino and antineutrino samples, important sources of systematics, such as the flux uncertainty, will not be correlated and these will dominate the sensitivity loss to CP violation in the large $\theta_{13}$ regime of interest in this study. In the proposals for all facilities, a near detector is envisioned so as to reduce the systematic errors to an ``acceptable'' level, which is usually considered to be around $5 \%$. The final reduction that can be achieved is, however, still a matter of ongoing debate among the experts. For this reason, in Fig.~\ref{fig:sys} we will vary the systematics to show how the optimization of the experiment for large $\theta_{13}$ changes with the level of systematic error ultimately achievable. All simulations made use of the GLoBES software \cite{Huber:2004ka,Huber:2007ji}. Ten years of data taking (with $10^7$ useful seconds per year) have been considered for all facilities. These 10 years have been divided in 2 for neutrinos $+8$ for antineutrinos at the SPL and $5+5$ for the T2HK setup. For the $\beta$-Beam, a $5+5$ configuration has been used for the standard setup, modified to $6+4$ at the higher energy and longer baseline $\beta$-Beam considered.

%These 10 years have been divided in 2 for neutrinos $+8$ for antineutrinos at the SPL and $5+5$ for the standard $\beta$-Beam and the T2HK setups modified to $6+4$ at the higher energy and longer baseline $\beta$-Beam considered.

%
\begin{figure}
\begin{center}
\includegraphics[width=.48\textwidth]{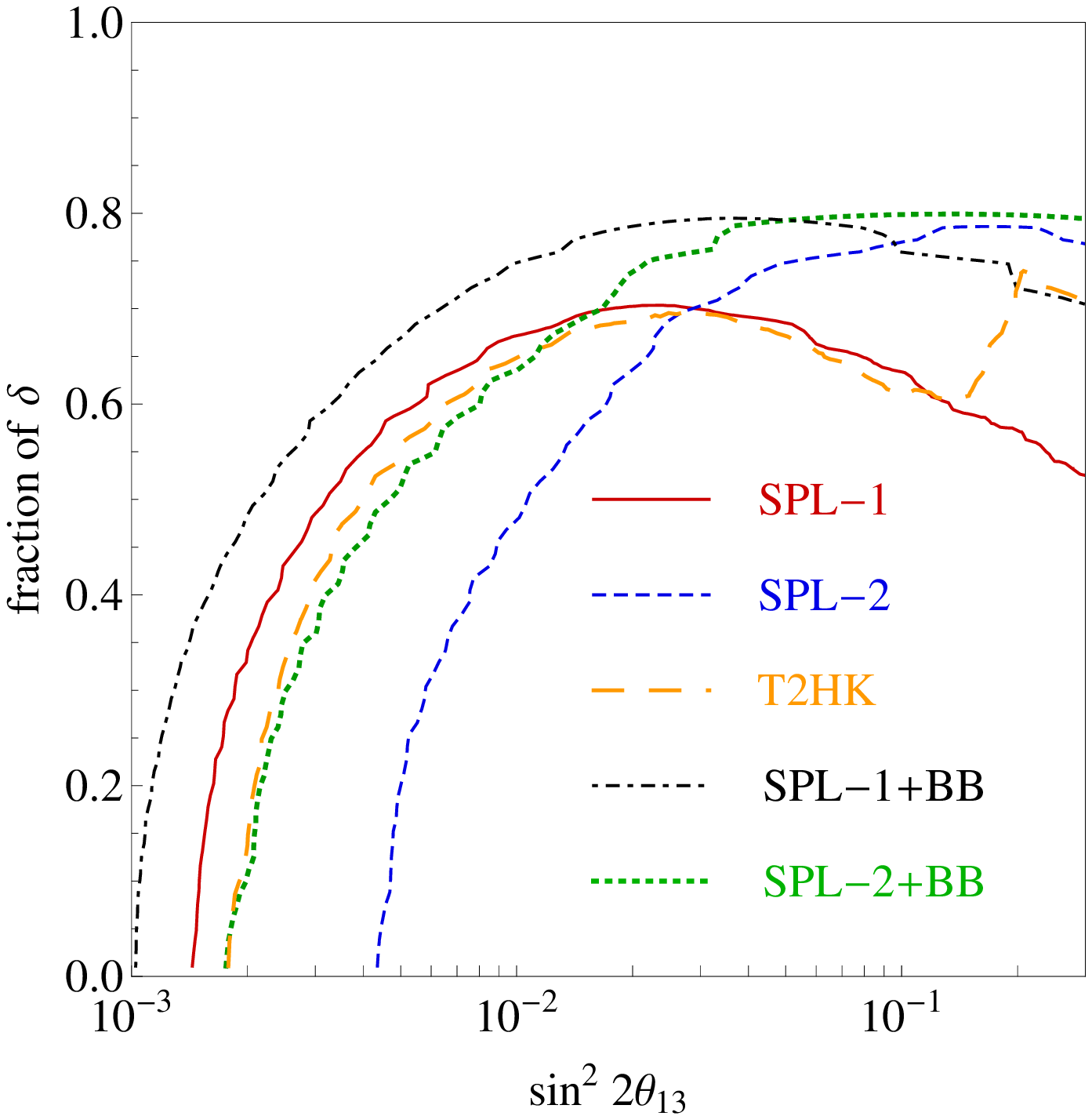}
\includegraphics[width=.48\textwidth]{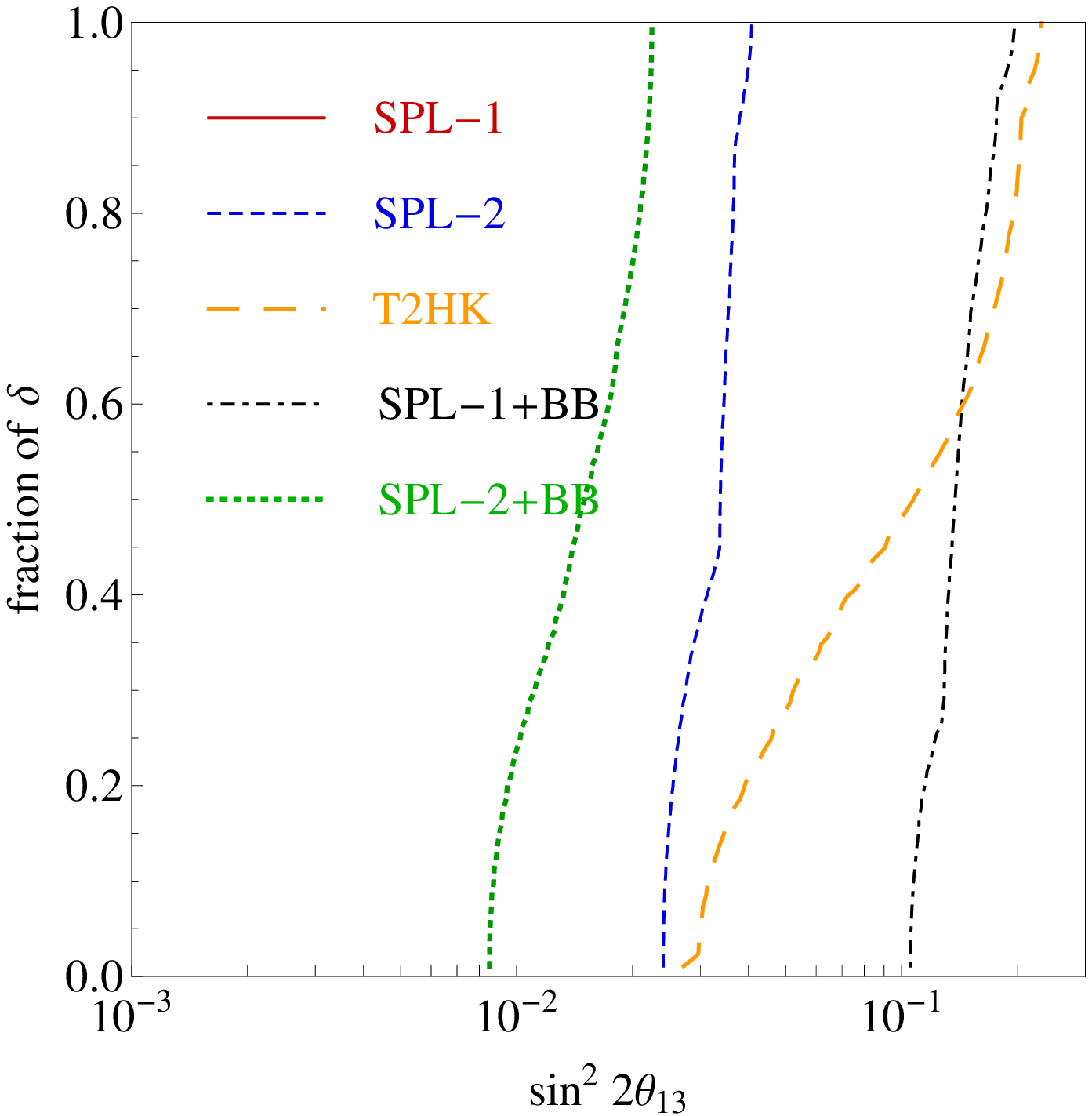} \\
\caption{Comparison of CP discovery potential (left panel) and mass hierarchy sensitivity (right panel) for several facilities. Solid red and dashed blue lines stand for the SPL Super-Beam at a 130 km (SPL-1) and 650 km baseline (SPL-2), respectively. T2HK is depicted by the orange longer dashed lines. Dot-dashed black lines stand for the $\gamma = 100$ $\beta$-Beam combined with the SPL at 130 km (SPL+BB-1) whereas dotted green lines depict the combination of a $\beta$-Beam with $\gamma = 250/100$ for $^{18}$Ne/$^8$Li with the SPL Super-Beam at 650 km (SPL+BB-2).}
\label{fig:sensitivities}
\end{center}
\end{figure}

In Fig.~\ref{fig:sensitivities} we compare the CP discovery potential (left panel) and the sensitivity to a normal hierarchy (right panel) for different setups. The CP discovery potential for a given value of $\theta_{13}$ is defined as the fraction of values of $\delta$ that, if realized by nature, would lead to a $3 \sigma$ exclusion of the CP-conserving values $\delta = 0$ or $\delta = \pi$ after marginalizing over all other free parameters. Similarly, the sensitivity to the normal mass hierarchy for a given value of $\theta_{13}$ is defined as the fraction of values of $\delta$ that would allow for a $3 \sigma$ exclusion of the inverted hierarchy hypothesis after marginalization over all free parameters. We have checked that the sensitivity to an inverted hierarchy is similar for the setups depicted here.

In the figure, the performance of the SPL Super-Beam aimed at 130 km baseline (CERN-Fr\'ejus baseline, dubbed SPL-1) is compared to the same setup aimed at a longer baseline of 650 km (CERN-Canfranc or CERN-Umbria baselines, SPL-2) and with the T2HK setup for reference. Both SPL-1 and the T2HK setups approximately match the first oscillation peak while the SPL-2 setup is very close to the second peak instead\footnote{A shorter baseline, around 450 km, or slightly higher energies would better match the second peak. We have simulated the 450 km option obtaining slightly better performance than for the 650 km one, particularly for small $\theta_{13}$, given the larger statistics available at the shorter baseline. The combination with the statistics-limited $\beta$-Beam also improves significantly at this better optimized, but less realistic, 450 km baseline.} (dashed blue lines). As can be seen, the measurement at the second oscillation peak makes the CP discovery potential improve in the large $\theta_{13}$ region currently favoured by global data. Moreover, this option would also guarantee a $3 \sigma$ measurement of the mass hierarchy down to $\sin^2 2 \theta_{13} > 0.04$, while the shorter 130 km baseline provides no sensitivity at all to this observable. For T2HK, a $3 \sigma$ measurement of the mass hierarchy is only guaranteed down to $\sin^2 2 \theta_{13} > 0.2$. On the other hand, the longer baseline implies a smaller event rate at the detector and thus, a smaller CP discovery potential for small values of $\theta_{13}$. 

We also show in the same figure the results for the Super-Beam combined with a $\beta$-Beam facility, for the two baselines under consideration. In vacuum, the $\nu_\mu \to \nu_e$ oscillation at the Super-Beam is related by CPT invariance to the $\beta$-Beam antineutrino channel ($\bar\nu_e \to \bar\nu_\mu$), which seems to imply little complementarity in this combination beyond increased statistics. However, matter effects prevent the exact CPT symmetry. Any difference between the oscillation probabilities at the two facilities can be attributed to matter effects and provides clean handles on the measurement of the mass hierarchy~\cite{Schwetz:2007py,Jansson:2007nm}. 
In the case of the shortest baseline, $L=130$ km, we show the standard combination of the SPL Super-Beam with the $\gamma=100$ $^{18}$Ne/$^6$He $\beta$-Beam, dubbed SPL-1+BB in Fig.~\ref{fig:sensitivities}. At this baseline, matter effects are very weak: the CPT symmetry is almost exact and thus the combination of facilities only provides sensitivity to the mass hierarchy around $\sin^2 2 \theta_{13} = 0.1$ (dot-dashed black lines). However, matter effects are large enough to slightly deteriorate the CP discovery potential through sign degeneracies, which appear for large $\theta_{13}$ and for positive (negative) values of $\delta$ for the Super-Beam ($\beta$-Beam). Hence, the combination of both facilities increases the CP discovery potential at large $\theta_{13}$, as can be seen in Fig.~\ref{fig:sensitivities}. 

We find that this combination, dubbed SPL-2+BB in the plot, also performs better at the longer baseline of 650 km, when the $^{18}$Ne/$^8$Li $\beta$-Beam oscillation is closer to the first oscillation peak while the SPL is near its second oscillation peak\footnote{We have also studied the results for this combination when $^6$He boosted at $\gamma = 150$ is used instead of $^8$Li. The results are practically identical in the region where $\sin^22\theta_{13}>0.04$, although slightly worse for very small values of $\theta_{13}$ (due to the smaller statistics at the detector).}. As already mentioned in the context of the wide band beam, the combination of information at the first and second oscillation maxima is a useful way of solving sign degeneracies, since these degenerate solutions appear in different places at either maxima~\cite{Diwan:2003bp,Ishitsuka:2005qi,Donini:2006dx}. While this combination turns out to be less effective for the wide band beam given the high neutral current contamination stemming from the high energy end of the spectrum tunned to the first peak~\cite{Huber:2010dx}, this issue is avoided by tunning different beams to the different maxima. Indeed, the sensitivity to the mass hierarchy is considerably improved, providing the best results among the setups studied here and ensuring a $3 \sigma$ measurement down to $\sin^2 2 \theta_{13} > 0.01$, regardless of the value of $\delta$. The CP discovery potential for large values of $\theta_{13}$ is also increased, reaching an $\sim 0.8$ CP fraction for $\sin^2 2 \theta_{13} > 0.03$. However, the net gain in the large $\theta_{13}$ region with respect to the performance of the SPL alone at the second oscillation peak probably does not justify the addition of the $\beta$-Beam companion.

\begin{figure}
\begin{center}
\includegraphics[width=.48\textwidth]{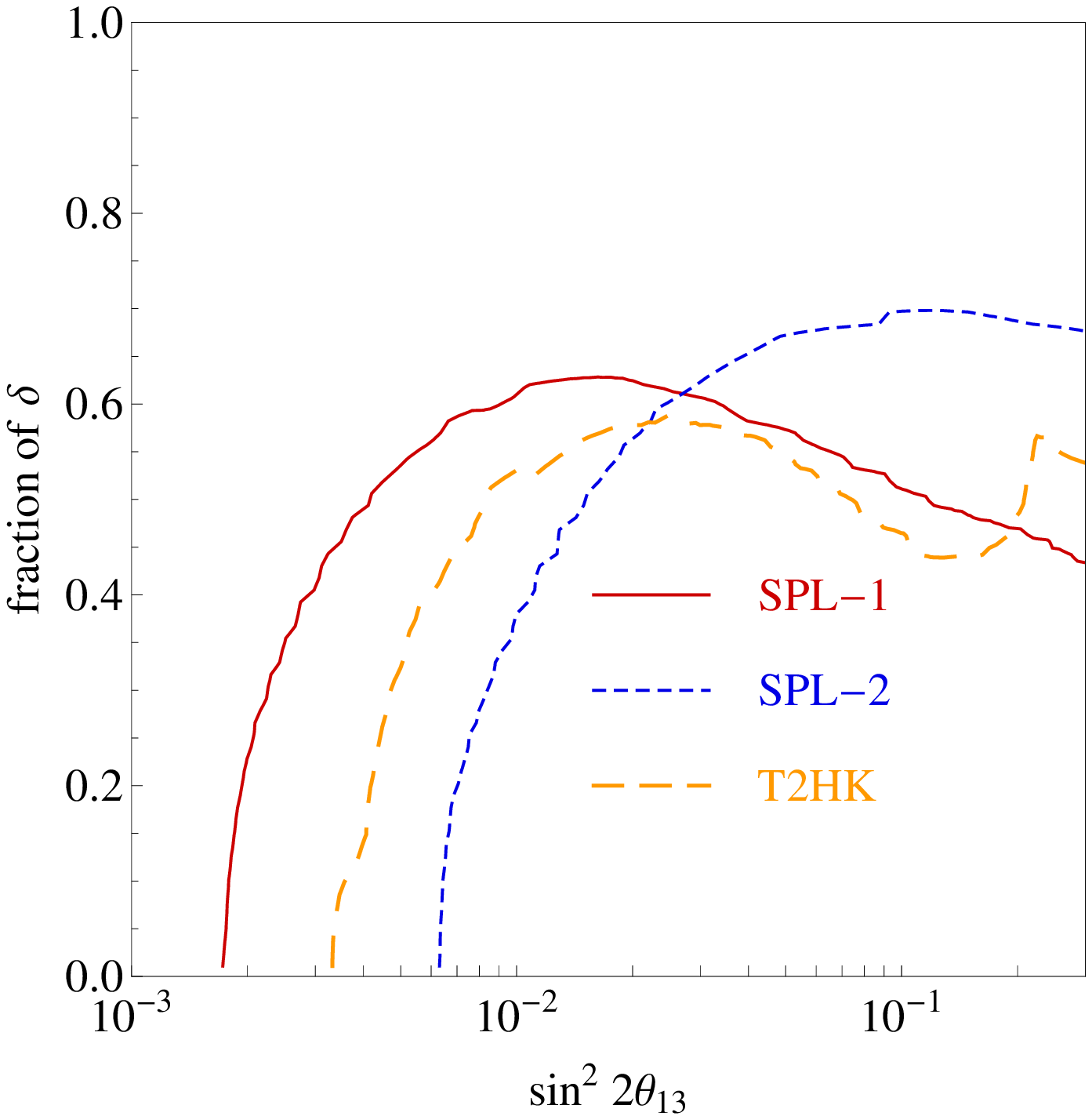}
\includegraphics[width=.48\textwidth]{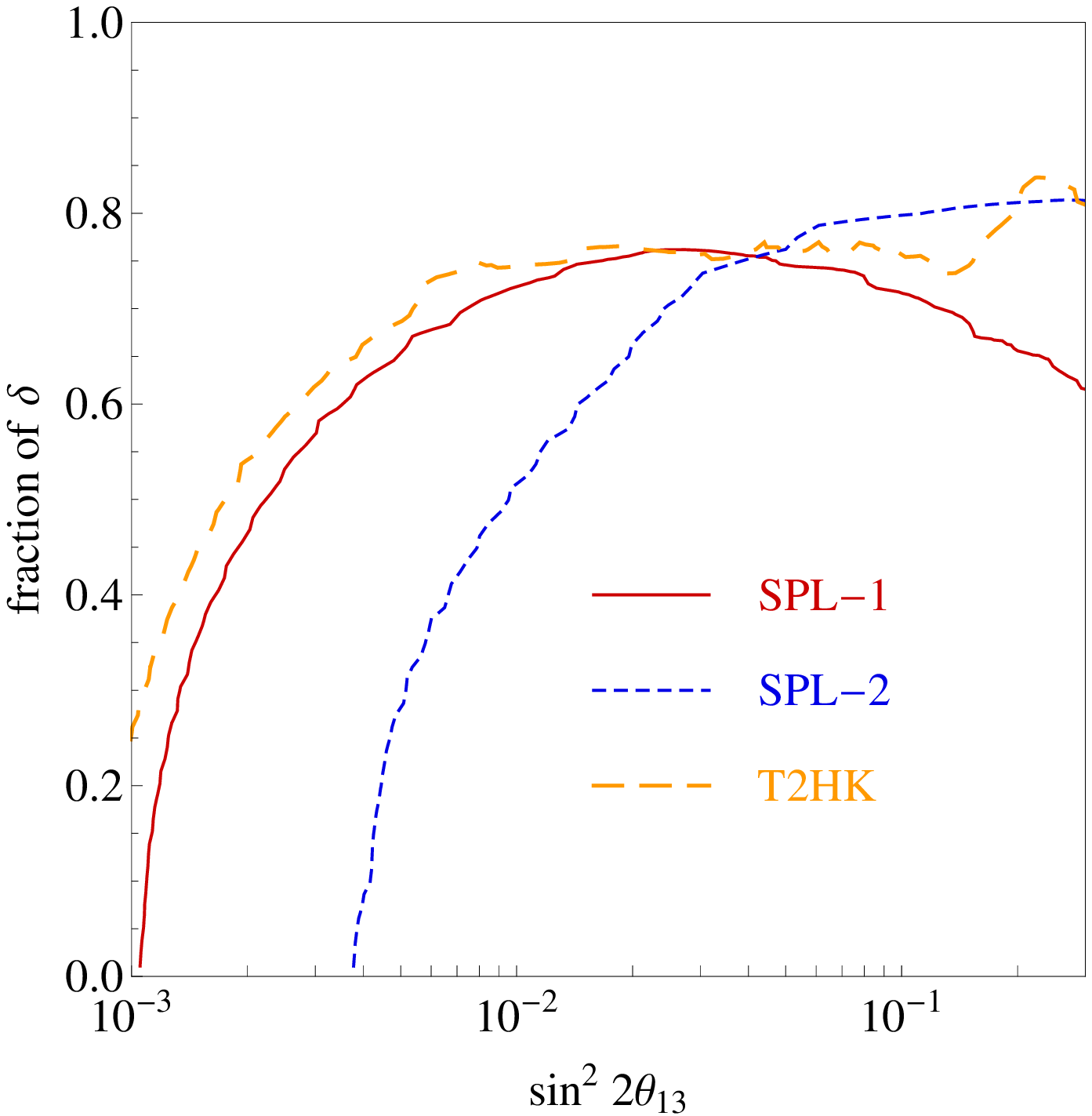} \\
\caption{Effect of the systematic errors on the CP discovery potential of the SPL-1 at 130 km (solid red lines) and the SPL-2 at 650 km (dashed blue lines). T2HK is depicted by the orange longer dashed lines. The left panel assumes ``large'' systematic errors of $10\%$ and $20 \%$ for the signal and background respectively while the right panel assumes ``small'' systematics of $2.5\%$ and $5 \%$ for signal and background. Fig.~\ref{fig:sensitivities} corresponds to an intermediate case of $5\%$ and $10 \%$ for signal and background.}
\label{fig:sys}
\end{center}
\end{figure}

A further advantage of performing the measurement at the second oscillation peak is that, since the CP interference term is leading in the probability, the CP discovery potential becomes less affected by systematic errors than measurements at the first oscillation peak. We show this effect in Fig.~\ref{fig:sys}, where the CP discovery potential for the SPL-1, SPL-2 and T2HK setups is compared. A pessimistic systematic of $10 \%$ in the signal and $20 \%$ in the background is assumed in the left pannel, while in the right panel optimistic values of $2.5 \%$ and $5 \%$ are considered instead. These can also be compared with the $5 \%$ and $10 \%$ assumed in Fig.~\ref{fig:sensitivities}. As can be seen, the performance at large $\theta_{13}$ for the SPL-1 and particularly of the T2HK setups are much more strongly affected by the size of the systematic error assumed than the longer baseline SPL-2 option. Indeed, for the most optimistic systematic scenario considered, the T2HK setup is comparable to (although slightly worse than) the SPL-2 one for large values of $\theta_{13}$, while SPL-2 is much preferable for large $\theta_{13}$ if the systematic errors are larger. 

To summarize, the setups at the first oscillation peak such as SPL-1 and T2HK would be preferable in order to maximize the CP discovery potential in the region where $\sin^2 2 \theta_{13} < 0.03$ or, maybe, in the case where the overall systematic errors on the signal are kept under control and around the $2.5\%$ level. On the other hand, the second oscillation peak offers a better chance to observe CP violation in the large $\theta_{13}$ region and it is in addition less affected by systematic errors. As for sensitivity to the mass hierarchy, the setups with $L=650$ km are always preferable due to the stronger matter effects at this longer baseline.

\section{Summary and conclusions}
\label{sec:summary}

The recent hint for large $\theta_{13}$ opens the window to the search for leptonic CP violation and the neutrino mass hierarchy at the next generation of neutrino oscillation facilities, thus bringing within reach the last unknowns in the picture of neutrino mixing. The optimization and design of these future facilities has usually followed a simple guideline: the best facilities are those providing sensitivity to $\theta_{13}$, CP violation and the mass hierarchy down to the smallest possible value of $\theta_{13}$. However, if the present hint for large $\theta_{13}$ is confirmed with increased statistics at T2K and by the reactor neutrino searches, the optimal facility for the next generation should instead focus in providing sensitivity to CP violation and the mass hierarchy for the largest possible fraction of values of $\delta$ in the range of $\theta_{13}$ favoured by data. 

We have argued that, with this change of paradigm in mind, a simple reoptimization of some proposed facilities for large $\theta_{13}$ is possible and desirable. Indeed, many of the next generation of neutrino facilities have been optimized so as to sit at the first oscillation maximum of the $\nu_e \to \nu_\mu$ oscillation (or its T conjugate channel). This choice maximizes the statistics at the detector and the ``atmospheric'' term in the oscillation probability, thus providing the best sensitivities for small $\theta_{13}$. The ``atmospheric'' term is CP-conserving and proportional to $\sin^2 2 \theta_{13}$. Hence, if $\theta_{13}$ is large, this term will tend to dominate the oscillation probability close to the first peak. Moreover, any systematic error in the signal, when applied to the leading contribution, can easily hide the subdominant CP-violating interference and will consequently deteriorate the CP discovery potential for large $\theta_{13}$. If the neutrino oscillation facility is instead placed at the second oscillation peak, the CP-violating interference has developed further and can constitute the dominant part of the oscillation probability. This setup can thus provide an increased CP discovery potential for larger values of $\theta_{13}$ and be less afflicted by systematic errors at the price of reduced statistics, that is, a poorer sensitivity for small values of $\theta_{13}$. Moreover, the increased baseline also enhances matter effects in neutrino oscillations and the sensitivity to the mass hierarchy is consequently improved. This option is therefore very attractive if the present preference for large $\theta_{13}$ is confirmed in the near future.

We have tested this idea with the SPL Super-Beam proposal. This setup has been optimized so that the neutrino energy is close to the first oscillation maximum when the neutrino beam is fired from CERN to a Mton Water \v{C}erenkov detector located 130 km away, in the Fr\'ejus underground laboratory. A further possible site for a large underground detector could be provided by the Canfranc or Umbria alternatives~\cite{Autiero:2007zj}, placed at $\sim 650$ km from CERN, which would bring the SPL beam close to the second peak. We have thus compared the performance of the two alternative sites for the SPL detector and with the T2HK proposal as a reference. As expected, the observation at the second peak provides increased CP discovery potential for large $\theta_{13}$ making it desirable over the first if $\sin^2 2 \theta_{13} > 0.03$. Moreover, the larger matter effects at this longer baseline also guarantee a $3 \sigma$ discovery of the mass hierarchy in the same range, which the Fr\'ejus option cannot probe by itself.     

The 650 km option studied also proved to be less dependent on systematic errors, as expected. For large $\theta_{13}$ the CP discovery potential of the shorter baseline of 130 km depends critically on the size of the systematic error and only becomes preferable to the 650 km option if the systematic error in the signal can be kept at the $2.5 \%$ level. For larger systematics, the 650 km option provides better CP discovery potential for large $\theta_{13}$ and always the best results for the sensitivity to the mass hierarchy.   

A companion $\gamma=100$ $\beta$-Beam facility is often envisaged together with the SPL Super-Beam at the 130 km baseline. This combination has been shown to also increase the CP discovery potential for large $\theta_{13}$. We find that this combination also performs better at the longer baseline. Since technical difficulties related to ion production and acceleration do not allow to reach the second oscillation peak with the $\beta$-Beam, we have tuned it as close as possible to the first peak at $L=650$ km. For $^{18}$Ne, we have assumed the maximum $\gamma$ achievable at the CERN SPS, which corresponds to $\gamma = 250$. We have also reduced the neutrino flux by a similar factor to take into account the boosted lifetime of the parent ions. However, the maximum value of $\gamma$ achievable at the SPS is not enough to bring the antineutrino flux produced from $^6$He decays to its first oscillation peak. Therefore, we have considered $^8$Li boosted to $\gamma=100$ instead. With this choice for the ions and boost factors, the $\beta$-Beam sits close to the first peak and offers complementary information to the measurements at the second peak of the Super-Beam. We find that the sensitivity to the mass hierarchy is significantly enhanced with respect to the standard SPL+$\beta$-Beam combination at the 130 km baseline, and the CP discovery potential is also increased in the large $\theta_{13}$ region with respect to the combination at $L=130$ km. As a final remark, it should be noted that the assumptions we have adopted for the $\beta$-Beam regarding the achievable ion decays per year, $\gamma$ factors and atmospheric backgrounds are rather conservative. Its performance in combination with the SPL would probably be considerably improved if any of these assumptions are relaxed. However, the gain provided by the addition of this second beam with respect to the performance of the SPL on its own is relatively small in the large $\theta_{13}$ and probably does not justify the need of a second beam.  

We believe that this simple optimization procedure for large $\theta_{13}$ is useful and desirable in view of present data and should definitely be considered if a large $\theta_{13}$ is established at a higher confidence level. While this optimization can in principle be considered for any type of facility, we find its mainly applicable to Super-Beams. Indeed, we tried a similar optimization for the $\beta$-Beam setup. However, the smaller neutrino flux combined with the longer baseline significantly reduced its performance. Moreover, the atmospheric neutrino background has been shown not to be negligible for the $\beta$-Beam and, while the signal is reduced at the longer baseline, the background remains at the same level. The situation is very different for Super-Beam experiments, where the main background source is beam-induced and thus reduced along with the signal when the baseline is increased. Notice, however, that the combined observation of the first and second oscillation peak at a wide band beam facility greatly deteriorates the usefulness of the second oscillation maximum data, given the high level of neutral current backgrounds which migrate to the lower energies from the part of the neutrino spectrum tunned to the first oscillation peak~\cite{Huber:2010dx}. Furthermore, short baselines are favored, given the lower statistics expected at the second maximum. Thus, we believe that the SPL facility with the longer baseline studied offers an almost optimal environment in which to exploit the sensitivity of the second oscillation peak to CP violation and the mass hierarchy for large $\theta_{13}$. Indeed, we have shown that this setup outperforms the standard SPL setup as well as T2HK for the $\theta_{13}$ range currently preferred by data.
As for the Neutrino Factory~\cite{Geer:1997iz,DeRujula:1998hd}, this optimization for large $\theta_{13}$ has already been performed by increasing $L/E$ with the low energy Neutrino Factory~\cite{Geer:2007kn,Bross:2007ts,Huber:2008yx,Tang:2009wp,FernandezMartinez:2010zza,Agarwalla:2010hk,Dighe:2011pa} alternative.

\begin{acknowledgments}

We would like to thank A.~Longhin for providing the SPL fluxes.

We acknowledge financial support from the European Community under the European Comission Framework Programme 7, Design
Study: EUROnu, Project Number 212372. The EU is not liable for any use that may be made of the information contained herein. 

PC has been supported by Comunidad Aut\'onoma de Madrid, and through projects FPA2009-09017 (DGI del MCyT, Spain) and HEPHACOS S2009/ESP-1473 (Comunidad Aut\'onoma de Madrid), as well as from the Spanish Government under the Consolider-Ingenio 2010 programme CUP, Canfranc Underground Physics, project number CSD00C-08-44022.

\end{acknowledgments}

\end{document}